# Nanostructured S@VACNTs Cathode with Lithium Sulfate Barrier Layer for Exceptionally Stable Cycling in Lithium-Sulfur Batteries

Mariam Ezzedine [1,z], Fatme Jardali [1], Ileana Florea [1,2], Costel-Sorin Cojocaru [1,z]

[1] Laboratoire de Physique des Interfaces et des Couches Minces (LPICM), CNRS, École Polytechnique, IP Paris, 91128, Palaiseau Cedex, France.
[2] Université Côte d'Azur, CRHEA, CNRS, 06903 Sophia-Antipolis, France.

[z]Corresponding Author E-mail Address mariam.ezzedine@polytechnique.edu, costel-sorin.cojocaru@polytechnique.edu

**Abstract**

Lithium-sulfur technology garners significant interest due to sulfur's higher specific capacity, cost-effectiveness and environmentally-friendly aspects. However, sulfur's insulating nature and poor cycle life hinder practical application. To address this, a simple modification to the traditional sulfur electrode configuration is implemented, aiming to achieve high capacity, long cycle life, and rapid charge rates. Binder-free sulfur cathode materials are developed using vertically aligned carbon nanotubes decorated with sulfur and a lithium sulfate barrier layer. The aligned CNT framework provides high conductivity for electron transportation and short lithium-ion pathways. Simultaneously, the sulfate barrier layer significantly suppresses the shuttle of polysulfides. The S@VACNTs with $Li_2SO_4$ coating exhibit an extremely stable reversible areal capacity of 0.9 mAh cm$^{-2}$ after 1600 cycles at 1C with a capacity retention of 80% after 1200 cycles, over three times higher than lithium iron phosphate cathodes cycled at the same rate. Considering safety concerns related to the formation of lithium dendrite, a full cell Si-Li-S is assembled, displaying good electrochemical performances for up to 100 cycles. The combination of advanced electrode architecture using 1D conductive scaffold with high-specific-capacity active material and the implementation of a novel strategy to suppress polysulfides drastically improves the stability and the performance of Li-S batteries.

**Introduction**

Within the realm of advanced electrode materials for lithium-ion batteries (LiBs), sulfur cathodes and silicon anodes are among the most attractive materials due to their excellent ability to store lithium ions and their abundance on Earth.[1,2] This has generated considerable interest and the potential for these materials to surpass the performance of commercially available LiBs.[3,4] Coupling of high-capacity electrodes in silicon-lithium-sulfur (Si-Li-S) and lithium-sulfur (Li-S) batteries offers considerably high theoretical specific energy of 1940 and 2600 Wh Kg$^{-1}$, respectively, significantly surpassing the 250 Wh Kg$^{-1}$ achievable with widely used traditional LiBs.[5–7] However, the implementation of S and Si electrodes faces challenges, as current prototype cells still suffer from both the electrodes' structural cracking associated with the large volume expansion during the lithiation and delithiation processes as well as from poor ionic and electronic conductivity.[8,9] Additionally, the Li-S battery chemistry encounters further obstacles. Indeed, the elemental sulfur ($S_8$) and the end-product lithium sulfide ($Li_2S$) are highly insulating materials.[10,11] During the charge and discharge processes, the active material is continuously lost due to the dissolution of high-order lithium polysulfides ($Li_2S_n$, $4 < n \leq 8$) in most of the organic liquid electrolytes, commonly known as the "shuttle effect".[12] This phenomenon not only results in ineffective sulfur utilization, restraining the practical specific capacity, but also leads to low coulombic efficiency and premature failure of the cells.[8,12]

Substantial progress has been achieved with input from the field of nanotechnology in order to overcome the described obstacles. On the cathode side, strategies such as combining sulfur with highly electrically conductive carbon material within electrodes (e.g., graphene,[13] micro/mesoporous carbon,[14,15] carbon nanofibers,[16,17] carbon nanotubes,[18,19] and hybrid structures[20]), utilizing polymer nanostructures[21,22], or developing new types of electrolytes [23,24] and protective coatings on the metallic lithium[25], have been proposed for enhancing the conductivity of sulfur and minimize the migration of polysulfides. Furthermore, recent studies have shown that the introduction of electrocatalysts, such as dual-atom Fe sites onto N-doped graphene[26] or nanocomposite electrocatalyst consisting of a carbon material and CoZn clusters[27], as well as the introduction of oxygen vacancies[28] can effectively mitigate the shuttle effect and enhance the performance of Li-S batteries. Similarly, on the anode side, significant efforts have been dedicated to improving cycling performance by engineering Si into nanostructures to circumvent the large volume expansion, including Si nanowires,[29,30] hollow Si nanostructures,[31,32] and porous Si.[33]

However, despite the high specific capacities of these active materials and the advancements to improve their performance, the architecture of the electrodes remains a crucial component for enhancing battery performance and has not changed since the introduction of LiBs. The design of the electrode requires the optimization of various parameters, such as particle size, porosity and thickness.[34] Conventional electrodes are traditionally produced through slurry casting with a thickness between 50 and 100 μm. Increasing the thickness of the electrode films means incorporating more active materials, thereby further increasing the energy density of the cells.[35] Unfortunately, one of the main limitations of slurry casting is that the thicker electrodes are prone to cracking and pulverizing from the current collector, leading to the disconnection of active materials. Apart from the delamination problem, the preparation of the slurry, composed of polymeric binder, a suspension of active material and conductive additives, faces challenges related to sluggish lithium-ion transport path through the highly tortuous electrode architecture.[36] To this end, altering the electrode structure, including electronic conductivity, active material distribution, and porosity, directly influences electrolyte infiltration, ionic transport, and the achievable balance of energy/power density in batteries. To obtain high-performance cells and to address the issues with each active material, long, thin and lightweight vertically aligned carbon nanotubes (VACNTs) can serve as conductive scaffold, effectively supporting sufficient active materials without experiencing delamination and guaranteeing better performance than the non-oriented structure.[37,38] VACNTs, grown directly on the current collector, offer excellent mechanical properties, ensuring the secure binding of electroactive particles. The intimate electrical contact between each particle of active material and the current collector establishes direct and fast pathways for electrons, reducing charge transfer resistances while maintaining the structural integrity. The large specific area CNT scaffolds can accommodate more than five times the sulfur species that traditional electrodes on conventional current collectors can handle and the high porosity of VACNTs contributes to enhanced accessibility for lithium-ions, promoting a high utilization of the sulfur active material.[39]

Here, we present a hierarchical electrode engineered via a bottom–up strategy that inherently exhibits high thickness and an abundance of pores in the structure of the carbon network that not only served as an excellent conducting agent but also can effectively trap the polysulfides in the electrode matrix, which in turn reduce the shuttle effect. Additionally, the addition of an uniform and ultrathin protective layer on the S@VACNTs using a solution of lithium sulfate ($Li_2SO_4$), drastically suppress the dissolution of polysulfides while allowing unhindered penetration of $Li^+$ ions.[40] This additional protection has been shown to be effective in increasing the lifespan and the cycling stability of the cells (**Figure 1**). To the best of our knowledge, $Li_2SO_4$ has not been reported as a protective layer, opening up a new approach in developing high-performance and high cyclability for Li-S batteries, thereby promoting the feasibility of Li-S batteries. To further exploit the versatility of these novel cathodes, a full cell was assembled, featuring a sulfur cathode vs silicon anode [38] with lithium source integrated into the silicon anode.

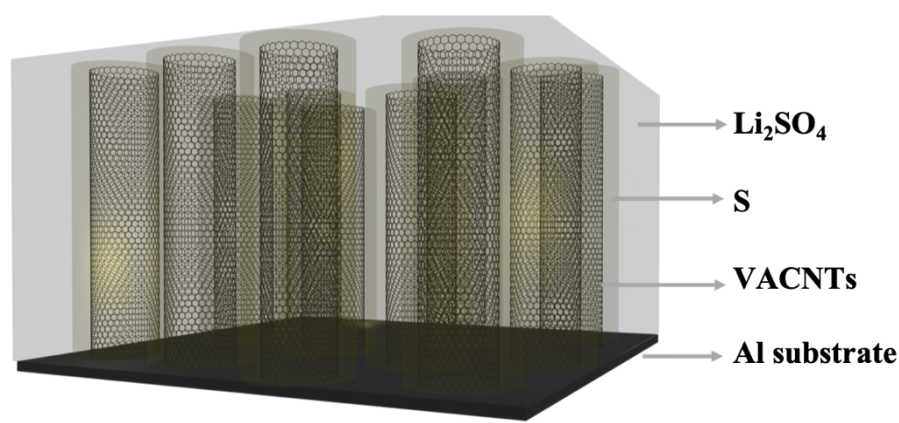

**Figure 1**. Schematic illustration of the hybrid S@VACNTs coated with $Li_2SO_4$ barrier layer electrode.

## Experimental

### Fabrication of the hybrid S@VACNTs electrode:

The VACNTs were synthetized on 50 μm thickness commercial aluminum (Al) foil, acting as a macroscopic current collector, using a quadruple hot-filament chemical vapor deposition reactor (4HF-CVD). The presence of four tungsten filaments (two at each gas inlet, methane ($CH_4$) and hydrogen ($H_2$), respectively, enables the growth of VACNTs with an increased degree of uniformity in the CNT's diameter and length and, most importantly, their synthesis on a large surface area.[41] To enable the VACNTs growth, a 30 nm aluminum oxide ($Al_2O_3$) film followed by a 3 nm thin layer of iron (Fe) were deposited on the conventional Al foil substrate by molecular beam evaporation (MBE) under high vacuum ($10^{-9}$ mbar) following similar procedure described in our previous work.[38,41] Fe catalyst pre-treatment under activated hydrogen was performed for 1.5 minutes at 600°C in the 4HF-CVD reactor. The $H_2$ gas was injected through the two dedicated tungsten filaments which were heated up to 2000°C (applied power 400 - 500 W) at a flow rate of 75 sccm and a reactor pressure of 8 mbar. Subsequently, the CNT growth was performed by exposing the pre-treated substrate to a gas mixture of $CH_4/H_2$ at flow rate of 50 and 20 sccm respectively and a reactor pressure of 12 mbar. The CNTs growth time of 30 minutes allowed for carpet height of about 40 µm. Subsequently, sulfur was deposited on the VACNTs carpet via a wet chemical deposition method. Sulfur solution was infiltrated into the VACNTs using a mixture of elemental sulfur powder dissolved in a solution based on carbon disulfide ($CS_2$) and isopropyl alcohol (IPA) in a (9:1) volume ratio. $CS_2$ is commonly used as a solvent due to its high solubility for S (~450 mg mL$^{-1}$ at 25 °C) and the addition of IPA into the solution enhances the infiltration of the solution into the pores by improving the wetting of the carbon surface.[42]

The impregnated S@VACNTs is subsequently dried in an oven at 40°C for 2 hours. The dried S@VACNTs is coated by a solution of $Li_2SO_4$ and IPA acting as a uniform and ultrathin protective layer to prevent the dissolution of polysulfides while allowing the penetration of Li$^+$ ions.

**Morphological, Chemical and Structural Characterization:**

The morphologies of the hybrid nanostructured electrodes were investigated with scanning electron microscopy (SEM) images using a HITACHI S 4800 FEG operating at 10 kV. Transmission electron microscopy (TEM) analyses and Energy-dispersive spectroscopy (EDS) analyses were carried out in the conventional TEM mode using a TITAN G2 electron microscope operating at 300 kV equipped with Oxford on side detector. Since the S@VACNTs nanostructures are very sensitive under the electron beam, for all TEM analyses a Gatan cryogenic sample holder was used in order to avoid any beam damage effects on the samples. GIXRD pattern were recorded with a Smartlab diffractometer using Cu−Kα radiation and scanned between 10° and 60° at a scan rate of 8°/min.

**Electrochemical Characterizations:**

For electrochemical tests, CR2032-type coin cells were assembled inside an argon filled glove box with S@VACNTs cathode, metallic lithium foil as anode, a combination of glass fiber (Whatman, GF/C) and propylene (Celgard, 2400) as separators soaked in the electrolyte solution. A fresh home-made electrolyte was prepared by dissolving 1 M bis(trifluoromethane)sulfonimide lithium salt (LiTFSI) and 0.25 M lithium nitrate ($LiNO_3$) in mixture of 1,3-dioxolane (DOL) and 1,2-dimethoxyethane (DME) (1:1 ratio, by volume). The electrolyte/sulfur ratio is 30 µL mg$^{-1}$. The electrochemical measurements were performed using a VMP3 (Bio-Logic) instrument at room temperature. All cells were kept idling for at least 6 hours at open circuit voltage before testing. The galvanostatic discharge/charge profiles of the cells were evaluated in a voltage range of 1.6-2.6 V vs. Li/Li$^+$. The specific capacity was calculated based on the mass of sulfur in the electrode and/or the electrode surface. The mass of VACNTs with a length of 40 µm is approximately 0.5 mg cm$^{-2}$. Cyclic voltammetry was evaluated for 10 cycles within a scan rate of 0.02 mv s$^{-1}$ between 1.6 and 2.6 V vs. Li/Li$^+$. The frequency range of the EIS measurements was 200 kHz to 1 mHz with an AC voltage amplitude of 10 mV at an open-circuit potential, after 1 and 10 cycles.

**Results and Discussion**

**Nanostructuring of the hierarchical hybrid electrode**

Scanning electron micrographs of pristine carbon nanotubes grown on ordinary aluminum (Al) foil with carpet thickness (CNTs length) of 40 μm are shown in Figure 2a. The CNTs appear well aligned along the normal axis, the carpet consisting of multi-walled (in average 3 to maximum 6 walls), dense ($10^{11}$ - $10^{12}$ tubes cm$^{-2}$), and tightly connected nanotubes to the Al substrate which is expected to provide a very good electrical conductivity in the nanostructured current collector. Figure 2b shows the scanning electron microscope (SEM) analysis of a VACNT carpet after S deposition by a wet chemical method. A closer view on the magnified images indicates a complete and uniform infiltration of S inside the CNT matrix and uniform coverage around the CNTs outer walls. The vertical structure of the CNT carpets remains unaffected even after S deposition. Furthermore, the S@VACNTs nanostructured electrode was characterized by transmission electron microscopy (TEM) and energy-dispersive X-ray spectroscopy (EDX). Due to the sensitivity of the sulfur under the electron beam, we used a cryogenic sample holder to avoid the beam damage during the observations. As shown in the low magnification illustrated in Figure 3a, b and c, the presence of a thin and homogeneous S layer can be identified on the side walls of the CNTs. A closer analysis of an area containing several S@VACNTs, as indicated by the green rectangle in the images illustrated in Figure 3b, evidences that the surface of nanotubes appears coarser for $CNT_2$ and $CNT_3$ compared to the pristine nanotube represented by $CNT_1$. TEM image of the pristine CNTs is shown in Figure S1. The sulfur presence was also evidenced by EDX spectroscopy on a large area (Figure 3d).

Figure S2 shows the grazing incidence X-ray diffraction (GIXRD) patterns of CNT after S deposition recorded at 2° angle of incidence. The 2θ scanned angle range was chosen from 10 to 60° with a step size of 0.02° and a counting time per step of about 1 second. The vertical red lines in the bottom axis correspond to the crystalline orthorhombic structure of sulfur (COD 9011362). The appearance of slight and broad peak centered around 2θ=24° corresponds to CNTs structure. The sharp peak located at 44.8° and the peak at 38.5° are assigned to the Al substrate (COD 9012429). The XRD analysis of sulfur exhibits a series of sharp and strong peaks with two prominent peaks at 2θ=22.9° and 27.6°, indicating a well-defined orthorhombic crystalline structure with space group Fddd, which are in excellent agreement with results reported in literature.[43,44]

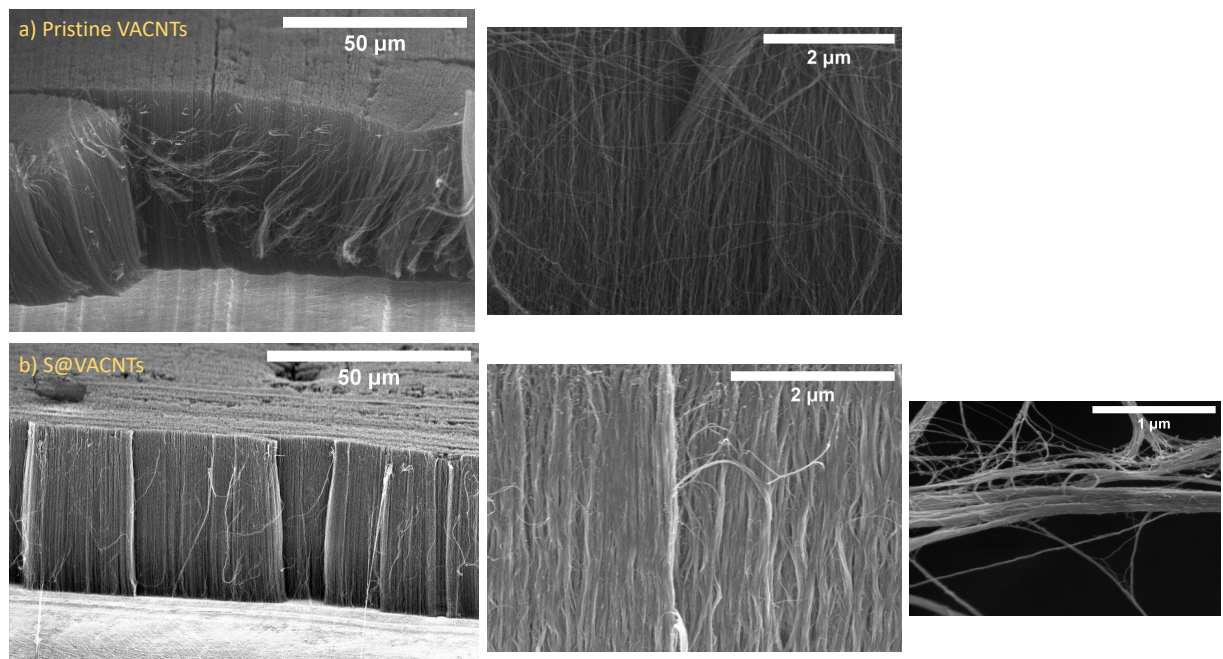

Figure 2. Cross section SEM micrographs and the corresponding high magnifications of pristine VACNT on Al foil (a) and S@VACNTs nanostructured electrode (b).

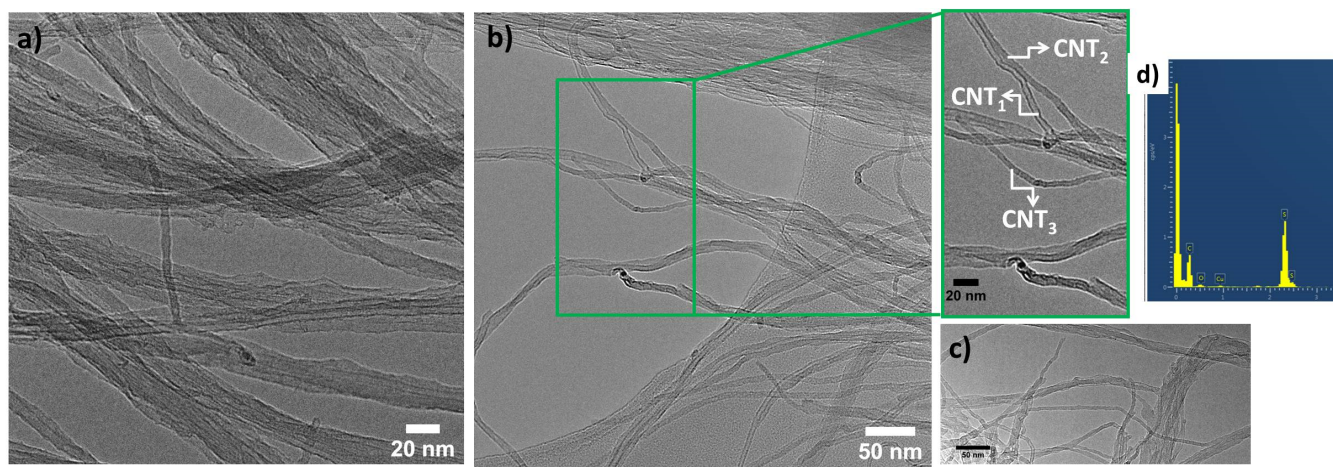

Figure 3. TEM images taken at different magnifications illustrating the morphology of S@VACNTs nanostructures (a), (b), (c). The zoom in the selected area in (b) illustrates three tubes, one uncovered $CNT_1$ and two covered $CNT_2$ and $CNT_3$; EDX spectrum taken in TEM mode on a large area containing S@VACNTs nanostructures (d).

**Electrochemical behavior of the nanostructured hierarchical hybrid electrode**

To eliminate any ambiguity in the terms, in the following, "lithiation" corresponds to the discharging process, and "delithiation" corresponds to the charging process in both Li-S batteries and Si-Li-S batteries.

The sulfur reduction is a multistep electrochemical process that involves various intermediate species following the reaction of $S_8$ with metallic lithium and the production of lithium polysulfides $Li_2S_n$ (2 < n < 8).[45] Figure 4a shows the cyclic voltammetry (CV) curves of the S@VACNTs electrode coated with $Li_2SO_4$ layer recorded at a scan rate of 0.02 mv s$^{-1}$ in the 1.6 to 2.6 V (vs. Li$^+$/Li) region for the 2$^{nd}$ and 10$^{th}$ cycle. The curves show two well-defined cathodic peaks at 2.3 V and 1.97 V corresponding to conversion of S$_8$ to soluble long-chain polysulfides (Li$_2$S$_n$, 4 < n < 8) following the reduction reaction of S$_8$ + 4Li$^+$ + 4e$^-$ → 2Li$_2$S$_4$ (A) and subsequently to the formation of insoluble short-chain polysulfides (Li$_2$S$_2$ and Li$_2$S) according to the reaction 2Li$_2$S$_4$ + 12Li$^+$ + 12e$^-$ → 8Li$_2$S (B), respectively. The oxidation peaks at 2.37 and 2.46 V in the anodic scan (C) correspond to the transformation of Li$_2$S into high-order lithium polysulfides and to S$_8$, respectively. The CV curves performed at different cycles thoroughly overlapped suggesting strong electrochemical reversibility and stability in the hierarchical hybrid VACNTs-based electrode.

Theoretically, the ratio of the areas under the two curves in region (B) and (A) is equal to 3 if the formed polysulfides are completely reduced to Li$_2$S.[13] A ratio of 2.6 is calculated by integrating the cathodic peaks in the 2$^{nd}$ cycle of Figure 4a, indicating that most lithium polysulfides were successfully converted to Li$_2$S$_2$/Li$_2$S. Achieving a ratio of 3 remains challenging due to the sluggish reaction kinetics in region (B). The area ratio of the cathodic (A and B) and the anodic (C) scans in the 2$^{nd}$ cycle shows a high Coulombic Efficiency of ∼ 96%.

The galvanostatic lithiation and delithiation profiles of the S@VACNTs nanostructured electrode coated with Li$_2$SO$_4$ barrier layer at 0.025 C rate are shown in Figure 4b. According to the voltage profiles at different cycles, two apparent lithiation plateaus and two delithiation plateaus are identified. The first lithiation plateau corresponds to the reduction of S$_8$ to high-order polysulfides occurring at about 2.4 V and the second lithiation plateau at 2.1 V corresponds to the further reduction of lithium polysulfides toward the highly insoluble and insulating solid-state lithium sulfides. During the delithiation process, reverse electrochemical reactions occur at around 2.3 and 2.4 V, respectively.

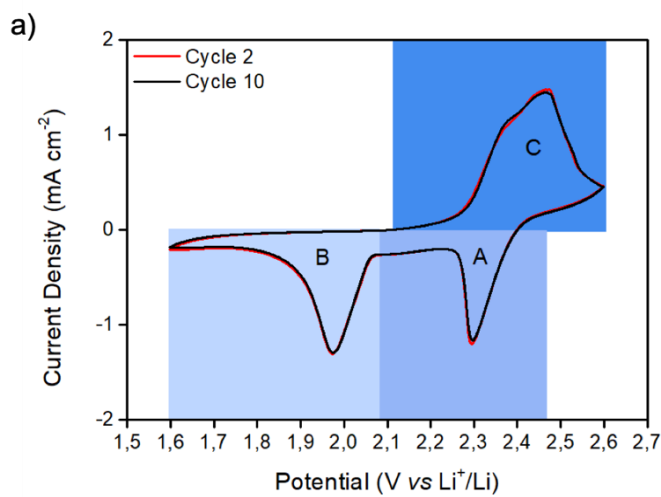

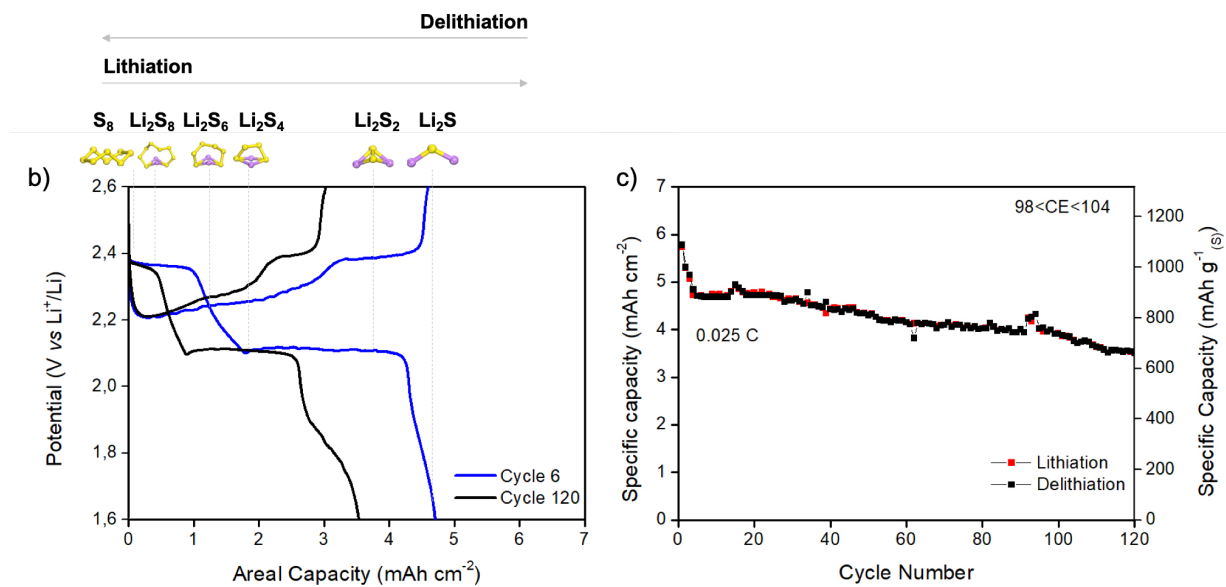

Figure 4. Electrochemical performances of S@VACNTs electrode coated with $Li_2SO_4$ layer: CV curve at a scan rate of 0.02 mV s$^{-1}$ (a). Galvanostatic discharge (lithiation)/ charge (delithiation) profiles at different cycles at 0.025 C rate, and the working mechanism of the Li-S battery at each stage (b). Areal and specific capacities during 120 cycles (c).

Figure 4c shows the areal and specific discharge/charge capacities at a rate of 0.025 C during 120 cycles within the potential window between 1.6 and 2.6 V. The specific capacities are calculated according to the mass of S with an average sulfur loading of 3.4 mg cm$^{-2}$. The S@VACNTs cathode exhibits a high initial areal discharge capacity of 5.7 mAh cm$^{-2}$ and a specific capacity of 1080 mAh g$^{-1}$ corresponding to a sulfur utilization of 64%. After the initial cycles, the S@VACNTs electrode displays electrochemical reversibility (as confirmed by the discharge/charge profiles for cycle 120 in Figure 4b) with an areal discharge capacity of 3.5 mAh cm$^{-2}$. Even after 120 cycles, the areal capacity of the electrode remains comparable to the areal capacity of commercial Li-ion batteries cathodes, which is approximately 4.0 mA h cm$^{-2}$.[39] The average Coulombic efficiency was calculated to be above 98%, demonstrating excellent cycling stability and high reversibility. The initial capacities for S@VACNTs electrodes with (Figures 4b and 4c) and without (Figure S3) the $Li_2SO_4$ barrier layer were 5.7 and 1.25 mAh cm$^{-2}$, respectively. After 10 cycles, the capacities decreased to 4.7 and 0.42 mAh cm$^{-2}$, representing capacity losses of 0.18% and 0.66%, respectively. Additionally, the areal capacity of the electrode with $Li_2SO_4$ is approximately four times higher than that of the electrode without $Li_2SO_4$. The introduction of $Li_2SO_4$ barrier layer, which have limited solubility in the electrolyte, showed a significant improvement not only in specific capacities but also in cycling stabilities, contributing to the retardation of the shuttle of sulfur-containing compounds (Figure S3).

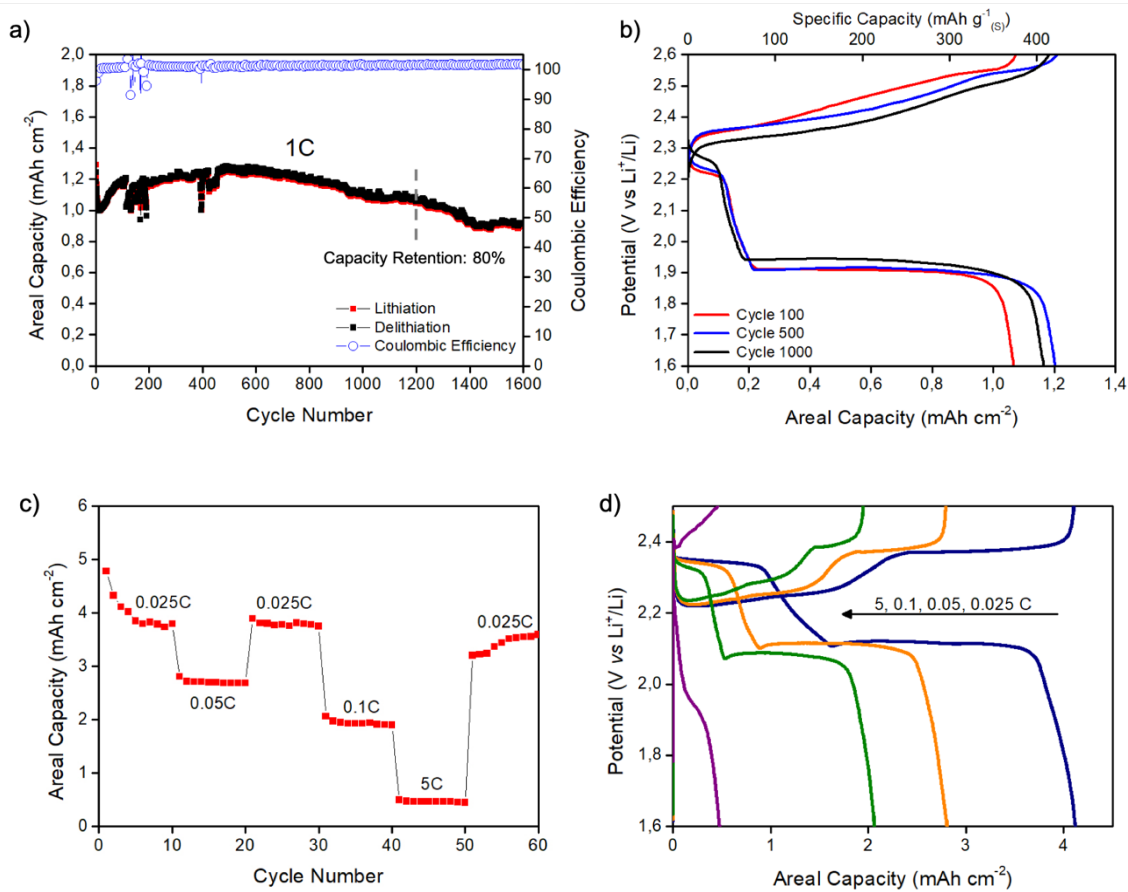

Figure 5. Long life cycling performance and Coulombic Efficiency of the S@VACNTs electrode coated with $Li_2SO_4$ layer at 1 C for 1600 cycles (a). Galvanostatic discharge/ charge profiles at 100, 500 and 1000 cycles at 1 C rate (b). Rate capability (c) and corresponding discharge/charge profiles (d) at different C rates.

The long-term cycling stability of the S@VACNTs electrode coated with $Li_2SO_4$ barrier layer with S loading of 2 mg cm$^{-2}$ was evaluated at 1 C for up to 1600 cycles, as shown in Figure 5a. The long-term cycling stability of the S@VACNTs electrode coated with Li2SO4 barrier layer with S loading of 2 mg cm-2 was evaluated at 1 C for up to 1600 cycles, as shown in Figure 5a. We observe a gradual increase in the specific capacities after the 12 cycles. This behavior is likely attributed to the slow kinetics reaction. During the initial cycles, sulfur undergoes an activation process, wherein a large electrode polarization leads to incomplete utilization of the active material at high rate. As cycling progresses, more lithium ions migrate at the electrode interface and infiltrate into the cathode's interior. Consequently, inactive sulfur on the internal surfaces becomes reutilized, resulting in stable capacities in subsequent cycles. Similar behavior has been reported in other sulfur-based cathodes.[46,47] The cell retained a reversible discharge capacity of 0.9 mAh cm$^{-2}$ after 1600 cycles, with an average Coulombic Efficiency of 99.99%, demonstrating outstanding cycling stability and high reversibility, significantly improving the electrochemical performance of Li-S batteries. The corresponding discharge/ charge profiles at 100, 500 and 1000 cycles are displayed in Figure 5b. Two obvious and stable discharge/ charge plateaus were retained even at high current density and after 1000 cycles revealing the high electrical conductivity and improved charge transfer kinetics through the electrode skeleton. The exceptional cycling properties can be attributed to the perfect adhesion between the CNTs (ensuring a direct connection to the macroscopic current collector) and the sulfur active material, as well as to the minimal tortuosity of the electrode structure. The constant plateaus corresponding to the generation of lithium polysulfides indicate that the highly porous structure of the VACNT carpet is effective in trapping the formed polysulfides and the $Li_2SO_4$ layer acts as a mechanical barrier to lithium polysulfides diffusion. Table S1 summarizes the areal and specific capacities obtained in this work in comparison to previously reported sulfur cathodes based on CNTs. The majority of sulfur electrodes [39,48–51] exhibit high specific capacities during the first cycle at low C rates. However, the performance declines by 50 to 56% during 40 cycles, 32% during 60 cycles, 33% during 100 cycles and 46% during 400 cycles compared to 31% during 1600 cycles obtained in our work. In comparison to these previous studies, our nanostructured electrode design, coupled with the novel approach of using lithium sulfate, demonstrates significantly superior high-stable cycling performance.

The rate capability of the hybrid S@VACNTs electrode coated with $Li_2SO_4$ barrier layer was investigated under various C rates, as shown in Figure 5c. The cathode exhibits a high reversible areal capacity of 4 mAh cm$^{-2}$ at 0.025C. When the C rate was increased to 0.05, 0.1, and 5, the cathode was able to deliver capacities of 2.6, 1.9, and 0.46 mAh cm$^{-2}$, respectively. Once the discharge/ charge rate returns to 0.025C after being cycled at high rates, a reversible areal capacity ∼ 3.6 mAh cm$^{-2}$ is recovered. This capacity is close to the one achieved during the initial cycles at 0.025C, indicating that the cathode structure remains unaltered even when subjected to high rates. Figure 5d gives the corresponding discharge/charge profiles. It is observed that the two plateaus are retained as the current rate increases from 0.1 C to 5 C, which again it is consistent with the high electrical conductivity and good charge transfer kinetics through the cathode and demonstrate the effectiveness of using the barrier layer. We also assessed the electrochemical contribution of the pristine VACNTs carpet and $Li_2SO_4$@VACNTs separately vs metallic lithium. We applied similar C rate conditions for both samples at 0.025 C. Figure S4 in the supporting information shows the potential profiles of 1$^{st}$ and 10$^{th}$ cycles vs capacity (mAh) for pristine VACNTs (a) and $Li_2SO_4$@VACNTs (b). The capacities for both samples are negligible, remaining in the μAh range.

To further understand the impact of the $Li_2SO_4$ protection layer in conjunction with the hybrid nanostructured electrode, electrochemical impedance spectroscopy (EIS) analyses were conducted to compare S@VACNTs cells with and without the $Li_2SO_4$ layer before and after cycling (Figure 6). Additionally, EIS analysis was performed on $Li_2SO_4$@VACNTs cell.

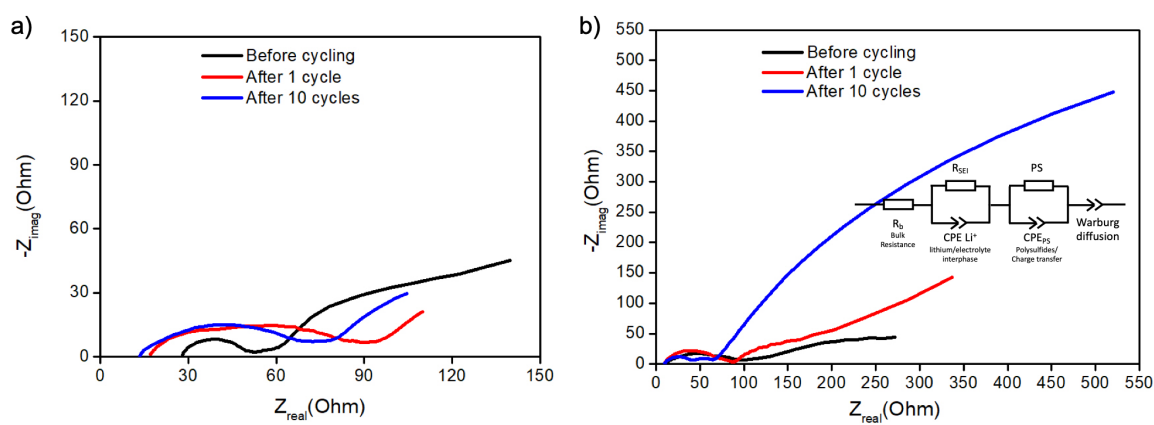

Figure 6. Nyquist plots before cycling and after 1 and 10 cycles of S@VACNTs coated with $Li_2SO_4$ (a) and S@VACNTs without $Li_2SO_4$ layer (b).

In order to interpret the Nyquist plots and gain a better understanding the contribution of the $Li_2SO_4$ layer, simplified equivalent circuits are proposed to fit the impedance spectra (inset in figure 6b). From high to low frequencies, the impedance spectrum comprises two semi-circles, one in the high-frequency region and another in the mid-frequency region (after cycling), followed by the Warburg impedance linked with lithium-ion diffusion in the low-frequency region. i) The bulk resistance ($R_b$) (such as the electrolyte, current collectors and separator) indicates the internal resistance of the cell at the intersection with the real axis. ii) The first semi-circle ($R_{sei}$) is associated to the lithium/electrolyte interphase (SEI). iii) The second semi-circle ($R_{PS}$) is related to the charge transfer reactions of sulfur intermediate species. iv) The straight line following the two semi-circles represents the Warburg impedance linked with lithium-ion diffusion. In Figure 6a, for S@VACNTs with the $Li_2SO_4$ layer, the $R_b$ is primarily associated with variations in electrolyte properties, such as chemical composition or viscosity. The $R_b$ decreased after cycling, a phenomenon already reported in Li-ion batteries.[52] This reduction can be attributed to a process that adjusts the internal components of a battery at the cycle's outset. This process involves the infiltration of electrolyte within the electrode, the distribution of electrode materials and the compact arrangement of the electrode structure, resulting in a decrease in bulk resistance. Alternatively, it can be explained that in the $Li_2SO_4$@VACNT cell, we also observe a decrease in $R_b$ after cycling (see Figure S5), indicating that $Li_2SO_4$ has an impact on lowering the internal resistance and improving the ionic conductivity in the cell. For S@VACNTs without $Li_2SO_4$ layer, the values of $R_b$ remain similar before and after cycling, suggesting no progressive loss of active material due to the irreversible precipitation of $Li_2S_2$/$Li_2S$. In middle-frequency region, the semi-circles show an increase with the characteristic frequency shifting to lower values (i.e. 3167 Hz before cycling, 159 Hz after 10 cycles) for the electrode coated with $Li_2SO_4$ layer. It can be attributed to the passivation of metallic lithium due to contact with the electrolyte, consuming lithium salt and solvents to form the SEI layer. In contrast, for the electrode without $Li_2SO_4$ layer, the first semi-circles shift to higher values of the characteristic frequency (i.e. 158 Hz before cycling, 3166 Hz after 10 cycles).[53] Such evolution can be linked to the chemical changes in the passivation layer, strongly influenced by the presence of the dissolved polysulfides that significantly alter its composition and thickness. Notably, no second semi-circle is visible for the cell with the $Li_2SO_4$ layer, highlighting the $Li_2SO_4$ layer's effectiveness in protecting high-order polysulfides from dissolving in the electrolyte during cycling. However, in the curves of the cell without the $Li_2SO_4$ protection layer, after cycling, a second semi-circle appears in the middle-frequency, associated with the depletion of polysulfides species. This leads to the precipitation of an insulating layer of solid $Li_2S_2$/$Li_2S$ deposited on the lithium anode surface.[54,55]

Nonetheless, some questions still linger and need further investigation. For instance, impedance measurements on symmetrical cells are needed in order to propose an overall interpretation of the different electrochemical reaction. Additionally, studying different potential cut-off during cycling is important in order to gain deeper understanding of the influence of the CNTs and of the $Li_2SO_4$ protective layer.

**Electrochemical behavior of the full Si-Li-S cell**

To highlight the potential of hierarchical hybrid S@VACNTs as a versatile cathode material for high-specific-energy rechargeable batteries and to mitigate the risks associated with the use of reactive metallic lithium as an anode material, which may lead to dendrite formation and pose severe safety concerns, we assembled a sulfur cathode and a silicon nanostructured anode.[38] To extend the lifespan of the full cell and to compensate for the loss of lithium through the formation of the SEI in the first cycles, a small lithium chip was integrated on top of the silicon electrode, creating direct contact with the current collector. During lithiation and delithiation processes, lithium ions shuttle back and forth between the electrodes. Due to its reducing properties, lithium ions will preferentially react with the Si nanoparticles rather than with the lithium chip.[7]

Si@VACNTs electrodes were characterized and tested in a half-cell configuration with metallic lithium counter electrodes in our previous work.[38] To investigate the capacity retention and the cell degradation, Si@VACNTs vs S@VACNTs coated with $Li_2SO_4$ layer cell was built with an optimized anode/cathode surface area ratio and the same electrolyte as in the Li-S cells. Figure 7 shows the electrochemical performances of Si@VACNTs vs S@VACNTs cell. The discharge/ charge voltage profiles are presented in Figure 7a for the 1$^{st}$ and 5$^{th}$ cycles, cycled between 1.3V-2.6V. The full cell has an initial lithiation/ delithiation capacities of about 6 mAh cm$^{-2}$ and 7.5 mAh cm$^{-2}$, respectively. The voltage profile presents the electrochemical reactions associated with the multi-step conversion process of the sulfur active material and an electrochemical process at ∼ 1.9 V corresponding to the 0-0.4V (*vs* Li/Li$^+$) delithiation potential of the Si electrode. The specific capacities display some fluctuations between the 10$^{th}$ and 80$^{th}$ cycles (Figure 7b), possibly attributed to the cracking of the SEI layer and to the activation of more Si atoms reacting with Li$^+$ ions. Despite a gradual capacity loss, the hybrid hetero-structures retained reversal capacities of about 1.2 mAh cm$^{-2}$ after 100 cycles. Even though nanostructured cathode and anode exhibit excellent and stable electrochemical performances when considered separately, the full cell experience faster capacities degradation. The unstable and poorer performance of the full cell can be attributed to the formation of a thick SEI layer upon cycling. During the delithiation process, the dissolved polysulfides are not oxidized back to elemental S, implying the consumption of Si and Li, thereby reducing the capacities of the electrodes. We notice that the Coulombic efficiency is higher than 100%. This mechanism is also observed in Li-S cells. Sulfur utilization gradually increases during the charge process as the sulfur is continuously exposed to the electrolyte. However, during the discharging process, only long chain polysulfides are formed, resulting in a lower discharging capacity than the charging capacity, and therefore the Coulombic efficiency is greater than 100%.

Nevertheless, we demonstrate the feasibility and cyclability of the hybrid architecture nanostructured electrodes. Notwithstanding the faster capacity fading as compared to the Li-S cells, the full cell Si-Li-S has a good electrochemical performance that can be attributed to the VACNTs carpets with a regular pore structure, highly ordered orientation, and an extremely large surface area. These characteristics result in significantly improved electrolyte accessibility and charge transport capability, which yield short and direct lithium-ion and electron pathways. Additionally, the VACNTs can alleviate the volume changes for both electrodes and adsorb the polysulfide intermediates due to the similar dimensions of the pores and polysulfide ions, thereby enhancing the cycling performance of the cells. Furthermore, the good electrochemical performance can also be attributed to the $Li_2SO_4$ barrier layer acting as mechanical barrier for preventing the diffusion of polysulfides. More studies will have to be conducted in order to identify the specific mechanisms that can explain why in a full Si-Li-S cell the stability is not as good as when considering each electrode separately.

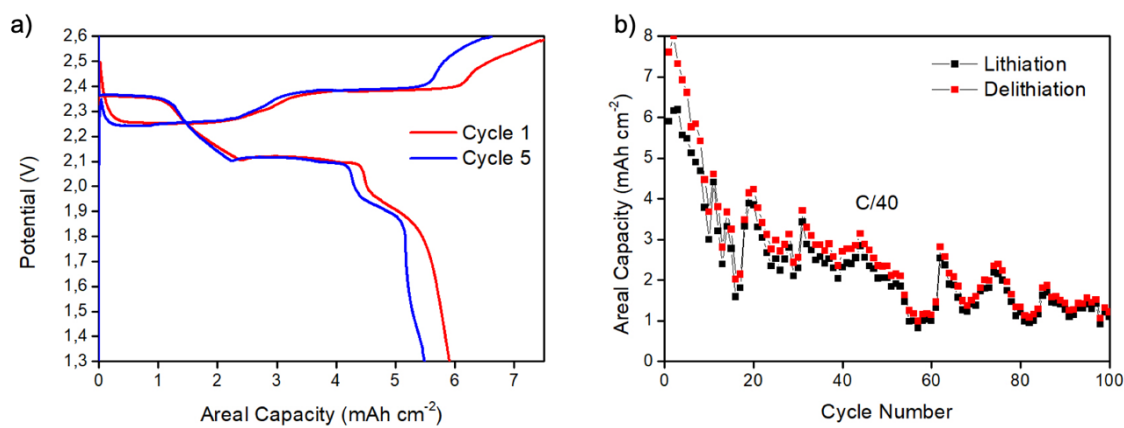

**Figure 7**. Lithiation/ delithiation profiles (a) and cycling performance over 100 cycles at a 0.025C of Si@VACNTS and S@VACNTs electrodes (b).

**Conclusions**

Hybrid hierarchical nanostructured S@VACNTs cathodes were directly synthesized on conventional Al foil current collectors through simple and easily scalable processes, eliminating the need for polymer binders or carbon additives. The aligned nanostructure of CNTs enhanced the electrolyte accessibility and the charge transport capability. The mechanically stable CNTs withstand the volume changes during cycling and preserves the structural integrity of the active materials throughout extended cycling. The introduction of a $Li_2SO_4$ barrier layer on the cathode, effectively suppressed the shuttle of polysulfides, leading to significantly improved electrochemical performances. After 1600 cycles at a 1C rate, the reversible retained capacities of 0.9 mAh cm$^{-2}$ are more than three times higher than those of lithium iron phosphate cathodes cycled at the same rate. The hierarchical hybrid nanostructure exhibited capacities twice and up to 7-fold higher than the theoretical specific capacities of commercial cathodes used in current LiBs. To assess the versatility of the 1D nanostructured architecture, we assembled a full Si-Li-S cell, demonstrating satisfying electrochemical performance over 100 cycles. A notable contribution of this study is to provide new insights for reducing the polysulfides diffusion, offering a promising direction for future development of the next-generation LiBs.

## Acknowledgments

The authors acknowledge financial support from: Chaire EDF "Energies Durable", the French state managed by the National Research Agency under the grant ANR-22-CE42-0030 (SPACESENSE) and the Investments for the Future program under the references ANR-10-EQPX-50 (pole NanoTEM) and ANR 2018-CE09-0033. The work was partially supported by Agence de l'Innovation de Défense – AID - via Centre Interdisciplinaire d'Etudes pour la Défense et la Sécurité – CIEDS - (project 2023 - HiPALis). The authors acknowledge the support of the X-ray crystallography facility, DIFFRAX, in Ecole Polytechnique, Institut Polytechnique de Paris. The authors thank Sandrine Tusseau-Nenez (PMC, Ecole Polytechnique, Institut Polytechnique de Paris) for her help for XRD data collection and their analyses. We would like to acknowledge the Centre Interdisciplinaire de Microscopie électronique de l'X (CIMEX). This work is part of the NanoMaDe-3E Initiative.